\documentstyle[aps,prb,multicol]{revtex} 
\input{epsf}
\newcommand{\beq}{\begin{equation}}  
\newcommand{\eeq}{\end{equation}}  
\newcommand{\bea}{\begin{eqnarray}}  
\newcommand{\eea}{\end{eqnarray}}  
  
\begin{document}     
\title{Critical statistics in a power--law random banded matrix ensemble}
\author{Imre Varga$^{a*}$ and Daniel Braun$^b$}
\address{$^a$Fachbereich Physik, Philipps Universit\"at    
	Marburg, D-35032 Marburg, Germany\\
	$^b$FB7, Universit\"at--Gesamthochschule Essen, D-45117 Essen,
	Germany}
\date{\today}     
\maketitle{}
\begin{abstract}
We investigate the statistical properties of the eigenvalues and eigenvectors
in a random matrix ensemble with $H_{ij}\sim |i-j|^{-\mu}$. It is known
that this model shows a localization--delocalization transition (LDT) as a 
function of the parameter $\mu$. The model is critical at $\mu=1$ and the 
eigenstates are multifractals. Based on numerical simulations we demonstrate 
that the spectral statistics at criticality differs from semi--Poisson 
statistics which is expected to be a general feature of systems exhibiting 
a LDT or `weak chaos'. 
\end{abstract}
\pacs{05.45.-a, 71.30.+h, 72.15.Rn}
\begin{multicols}{2}
\narrowtext
In a recent paper Bogomolny {\it et al.} \cite{Bogom} have investigated 
a number of dynamical systems and found remarkable similarities between 
the spectral statistics of pseudointegrable billiards and the critical 
statistics in the Anderson model at the mobility edge. The latter
has been investigated numerically by a number of authors 
\cite{Shkl,GOE,GUE,GSE,Daniel,IV}. At the metal--insulator transition 
(MIT) properties intermediate between those predicted by random matrix 
theory (RMT) \cite{Mehta,PhRep} and those of uncorrelated spectra with 
Poisson statistics were found. On one hand the spectral statistics for 
small differences in energy is reminiscent of the universal level 
repulsion in metals, i.e. the nearest neighbor level
spacing distribution $P(s)$ behaves as $P(s)\sim s^{\beta}$ for $s\to 0$,
with $\beta=1$, 2, or 4 for orthogonal, unitary, or symplectic symmetry 
of the system. On the other hand for $s\gg 1$, $\ln P(s)\sim -as$ with 
a constant $a$ depending weakly on $\beta$ but strongly on the 
dimensionality $d$, which is reminiscent of uncorrelated energy spectra in
Anderson insulators. These results were found in numerical simulations on 
$d$--dimensional cubic systems using periodic boundary conditions (BC). 
Recently Braun {\it et al.} \cite{Daniel} discovered  that the shape of 
$P(s)$ depends strongly on the choice of the BCs. Upon averaging over BCs 
they concluded that for orthogonal symmetry $P(s)$ is very close to the 
semi--Poissonian form
\beq
   P(s)=4s\,e^{-2s}\,.
\label{semiP}
\eeq
The energy level statistics is universal at the MIT in the sense that the
level number variance $\Sigma^2(\bar{N})=\langle (\delta N)^2\rangle$ is
proportional to the mean number of levels $\bar{N}\gg 1$ and the coefficient
$\chi$ is independent of the BCs. Dependence of the $P(s)$ function on the
BCs in $d=2$ quantum--Hall systems \cite{Schw} and appearance of the 
semi--Poisson statistics for a $d=2$ Anderson model with symplectic 
symmetry \cite{Evang} obtained as an average over the BCs have also been 
reported since then. Interestingly similar spectral statistics has been
obtained for the case of two interacting particles in a one dimensional 
disordered system at the interaction strength producing maximal mixing of
the noninteracting basis.
\par
At the transition point (MIT) the statistical properties of the spectra 
and of the eigenstates are linked to one another. A remarkable 
relation between the level compressibility $\chi$ and the density correlation
dimension of the eigenstates $D_2$ has been derived in \cite{Chalker},
\beq
   \chi=\frac{1}{2}\left (1-\frac{D_2}{d}\right )\,.
\label{lcomp}
\eeq
The dimension $D_2$ describes how the fourth moment of the components 
of an eigenfunction $\psi$ scales with the linear length of the system 
\cite{Martin}. Generally, the $2p$-th moment scales as
$\sum_i\langle |\psi_i|^{2p}\rangle\propto L^{-D_p(p-1)},$
where in the case of multifractality $D_p$ is a nonlinear function of $p$.
\par
The same parameter $D_2$ describes the scaling of the probability overlap 
of two states with an energy separation substantially exceeding the mean 
level spacing; hence the name of density correlation dimension.
It has been shown \cite{FM} that the heavily fluctuating local densities 
still produce a considerable overlap so that at the MIT level repulsion 
is still present.
\par
In Ref. \cite{Bogom} quantum chaotic systems were numerically compared 
to one of the simplest models that provides semi--Poissonian statistics: 
the short--range plasma model (SRPM). The model describes $N$ levels that 
repel each other logarithmically as in conventional RMT \cite{Mehta}, but 
with the interaction restricted to nearest neighbors only. For this SRPM 
many quantities can be computed analytically in the large $N$ limit. The 
spacing distribution is given by Eq.~(\ref{semiP}), the two--level 
correlation function reads
\beq
   R(s)=1-e^{-4s}\,.
\label{SRPM-rs}
\eeq
The corresponding level number variance,
\beq
   \Sigma^2(L)=\frac{L}{2}+\frac{1}{8}(1-e^{-4L})\,
\label{SRPM-s2}
\eeq
leads to the level compressibility $\chi=1/2$ which is also the upper
limit assumed by the right hand side of Eq. (\ref{lcomp}). Another way
of obtaining semi--Poissonian statistics is to leave out every other
element from an otherwise uncorrelated sequence of levels. Such a 'daisy'
model has been studied recently in Ref.~\cite{tSel}.
\par
By allowing the logarithmic repulsion to act without limits in the SRPM one 
recovers the RMT result in which the level spacing distribution is well 
approximated by the Wigner--surmise
\beq
   P(s)=\frac{\pi}{2}s\,e^{-\frac{\pi}{4}s^2}\,,
\label{RMT-ps}
\eeq
and the the two--level correlation function is given by \cite{PhRep}
\beq
   R(s)=1-c^2(s)-\frac{dc(s)}{ds}\int_s^{\infty}c(s')ds'\label{RMT-rs}
\eeq
with $c(s)=\sin(\pi s)/\pi s$. From this correlation function the
level number variance of standard RMT (up to $1/L$ corrections) follows
\beq
   \Sigma^2(L)=\frac{2}{\pi^2}
   \left [\ln (2\pi L)+1+\gamma - \frac{\pi^2}{8}\right ]\label{RMT-s2}\,
\eeq
where $\gamma=0.5772...$ denotes Euler's constant. The level compressibility 
vanishes; the levels can be thought of as particles of an incompressible 
fluid \cite{Mehta,PhRep}.
\par
In the present paper we investigate the critical spectral statistics and
the multifractalilty of the eigenstates in a random matrix model originally
proposed by Mirlin {\it et al.} \cite{Mirlin} and later discussed in
\cite{Kravt,Krav-loc,Mirlin-rev,Evers}. The $N\times N$ matrices in this 
model are real symmetric and all entries are drawn from a normal distribution 
with zero mean, $\langle H_{ij}\rangle=0$. The variance depends on the 
distance of the matrix element from the diagonal \cite{Mirlin,Kravt},
\beq
   \langle (H_{ij})^2\rangle=[1+(|i-j|/B)^{2\mu}]^{-1}\,.
\label{matrix}
\eeq
In Ref.~\cite{Mirlin} it has been shown using field theoretical methods 
that for a fixed $B\gg 1$ the statistical properties of such
matrices for $\mu<1$ resemble those of RMT. On the other hand 
values $\mu>1$ in the limit $N\to\infty$ lead to uncorrelated
spectra similarly as in the case of banded random matrices \cite{FM-BRM}.
The case $\mu=1$ was proven to be of special importance for it
produces critical (multifractal) eigenstates and critical statistics. 
Due to the simplicity of the basic model and the possibility of 
analytical treatment of a MIT further details have been revealed recently
by Kravtsov \cite{Krav-loc} and Mirlin \cite{Mirlin-rev}. 
\par
Here we compare the statistical properties of energy spectra and
eigenfunction properties of this random matrix model at $\mu=1$ with the 
semi--Poisson statistics of the SRPM in a regime so far inaccesable to 
field theoretical methods, at $B=1$.
We study the shape of the nearest neighbor spacing distribution $P(s)$, the
two--level correlation function $R(s)$ and obtain the spectral comressibility
$\chi$ from the asymptotic behavior of the number variance $\Sigma^2$.
We also show the multifractality of the eigenstates and provide the
correlation dimension $D_2$. We will show that for $B=1$ relation 
(\ref{lcomp}) is satisfied provided that the spectra and the eigenstates
are limited to a small portion around the middle of the band.
\par
In our numerical investigation we have collected the spectra of $N\times N$ 
matrices for $N=800\dots 4800$. The power law nature of 
the problem did not allow the use of efficient algorithms (e.g. the 
Lanczos--algorithm) usually applied for the study of the MIT in the Anderson 
model. We have performed a very careful unfolding since the $N\to\infty$ and
$L\to\infty$ limits are approached very slowly. Based on the 
values of $N$ we considered, we were able to extrapolate the expected 
behavior of several quanities in the $N\to\infty$ limit. In all of our 
subsequent discussion we limit ourselves to the middle half of the spectra
both for the eigenvalues and for the eigenvectors.
\begin{figure}[tbh]
\epsfxsize=8cm
\epsffile{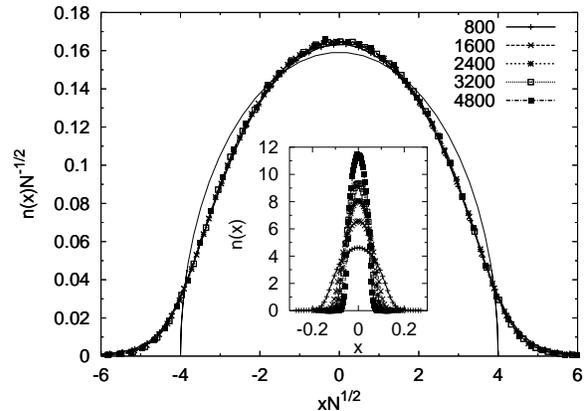}
\vglue 0.2cm
   \caption[DOS]{Density of states of the random matrix
   ensemble at $\mu=1$ and $B=1$ rescaled with $\sqrt{N}$. The continuous
   line is the semi--circle law $n(x)=\sqrt{1-(x/4)^2}/2\pi$. The inset
   shows the density of states before scaling.}
\label{dos-fig}  
\end{figure}
\begin{figure}[tbh]
\epsfxsize=8cm
\epsffile{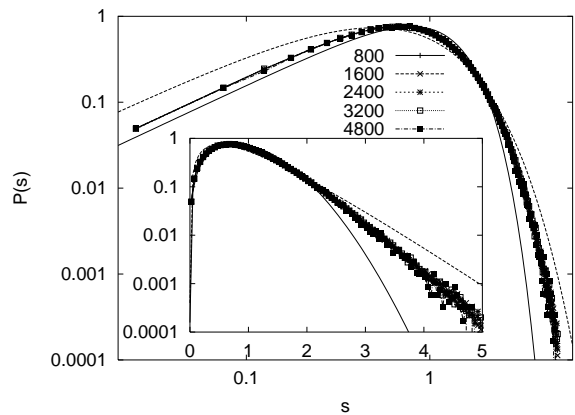}
\vglue 0.2cm
   \caption[PS]{Level spacing distribution $P(s)$ on a log-log scale. 
   The continuous line 
   shows the Wigner--surmize (\protect\ref{RMT-ps}) and the dashed line 
   represents the semi--Poisson distribution (\protect\ref{semiP}). The 
   insert shows the same functions on a lin-log scale.}
\label{ps-fig}  
\end{figure}
First the density of states of the model is plotted in Fig.~\ref{dos-fig}. The 
rescaled function is different from the semi--circle law. 
In Fig.~\ref{ps-fig} we show that the $P(s)$ is independent of $N$ and shows
a non--negligible deviation from the semi--Poissonian (\ref{semiP}) form. 
The deviation is seen both for the low--$s$ and the large--$s$ part of the 
function. The low--$s$ part of the $P(s)$ is of the form $P(s)\sim s$, 
but with a different prefactor. The large--$s$ behavior differs considerably 
from the $\ln P(s)\sim -2s$ form expected from the semi--Poisson 
distribution (\ref{semiP}). The deviation from the semi--Poisson statistics 
seems to persist in the $N\to\infty$ limit. The `peakedness' 
parameter \cite{IV} ($q=\langle s^2\rangle^{-1}$) of the $P(s)$ converges 
to a value $q=0.7342\pm 0.0009$ larger than that calculated for 
the semi--Poisson distribution (\ref{semiP}) for which $q=2/3$. We see that 
the shape of the $P(s)$ is indeed intermediate between the semi--Poisson and 
the Wigner--surmise. The spacing distribution in a similar random matrix
model has been studied by Nishigaki \cite{Nishigaki} whose result applied 
for our case is $q=0.7624$ which is close to our value.
\par
\begin{figure}[tbh]
\epsfxsize=8cm
\epsffile{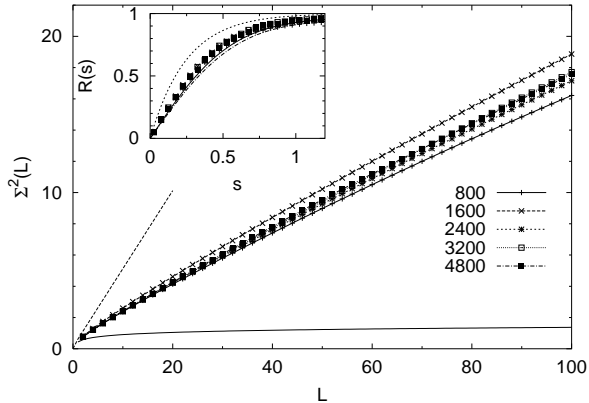}
\vglue 0.2cm
   \caption[s2]{Level number variance $\Sigma^2(\bar{N})$ as a function of 
   the unfolded $\bar{N}=L$. The continuous line shows the RMT case, Eq.
   (\protect\ref{RMT-s2}), and the dashed line 
   represents a semi--Poisson behavior of the form
   Eq.~(\protect\ref{SRPM-s2}). In the insert we compare the two--level 
   correlation function with the one obtained from the SRPM 
   Eq.~(\protect\ref{SRPM-rs}) (dashed line). The data are a little bit 
   better descirbed by the function obtained by Kravtsov 
   \protect\cite{Krav-loc} (solid curve). The RMT function, 
   Eq.~(\protect\ref{RMT-rs}), is given as a dashed-dotted curve.}
\label{s2-fig}  
\end{figure}
\begin{figure}[tbh]
\epsfxsize=8cm
\epsffile{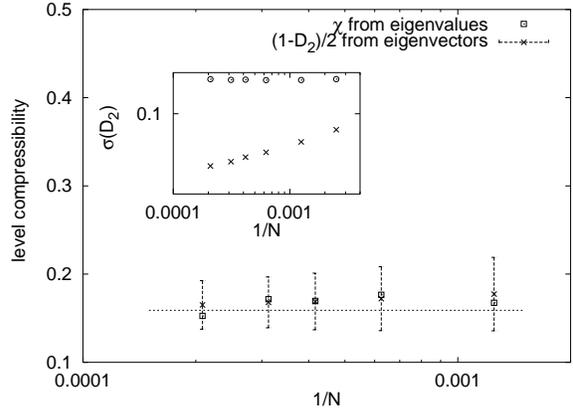}
\vglue 0.2cm
   \caption[chi]{Level compressibility as a function of the logarithm of
   inverse matrix size, calculated from the eigenvalue spectra 
   (open squares) and from the multifractal properties of the eigenstates 
   (crosses) [c.f. Eq.~(\protect\ref{lcomp})]. The dashed horizontal line
   corresponds to the theoretical estimate of $\chi=1/(2\pi)=0.159154\dots$.
   The errorbars are calculated from the variance of the correlation
   dimension, $\sigma (D_2)$, the errorbar of the data from the spectrum is
   smaller than the size of the symbol.
   The insert shows the $N$--dependence of $\sigma (D_2)$ if the averaging 
   is over the full band (circles) or the middle half of the band (crosses).}
\label{chi-q}
\end{figure}

Fig.~\ref{s2-fig} shows that the level number variance $\Sigma^2(L)$ converges
to a function that has a linear part, $\Sigma^2(L)\sim\chi L$ with a slope 
less than unity, $\chi< 1$, showing a nonzero compressibility of the spectra. 
However, this compressibility seems to converge to a value smaller than the 
$\chi=0.5$ of the SRPM. This behavior is reflected also in the two--level 
correlation function that shows deviations from Eq.~(\ref{SRPM-rs}), while
it is better described by the function calculated in 
\cite{Krav-loc,Mirlin-rev}. It is of the form of (\ref{RMT-rs}) with
$c(s)=0.25\sin (\pi s)/\sinh (\pi s/4)$. The same kernel applies
also for models discussed in Refs.~\cite{Nishigaki,Mutt}.
We have checked the validity of the relation (\ref{lcomp}) by comparing the
level compressibility obtained from the energy spectra with the value extracted
from the multifractal dimension $D_2$ of the eigenstates. The results are
shown in Fig.\ref{chi-q}. As $N\to\infty$ the data extrapolate nicely to a 
common value of $\chi=0.169\pm 0.019$ that is less than the value 0.5 
expected from the SRPM and is also smaller than the value $0.27$ found at 
the Anderson transition~\cite{Daniel}. Our result is in full aggreement with
the recent analytical estimate of $\chi=1/(2\pi)$ obtained in 
Refs.~\cite{Krav-loc,Mirlin-rev} and also earlier in an analogous random
matrix model~\cite{Blecken}. Note that the $D_2$ value is obtained as an
average of the states in the middle of the band. The inset of Fig.\ref{chi-q}
shows that indeed the variance of $D_2$ decreases with increasing $N$ while
taking the full set of eigenstates results in a distribution of these
exponents that remains broad even in the $N\to\infty$ limit a phenomenon
already noticed in \cite{Shober} and discussed in \cite{Evers}.
\begin{figure}[tbh]
\epsfxsize=8cm
\epsffile{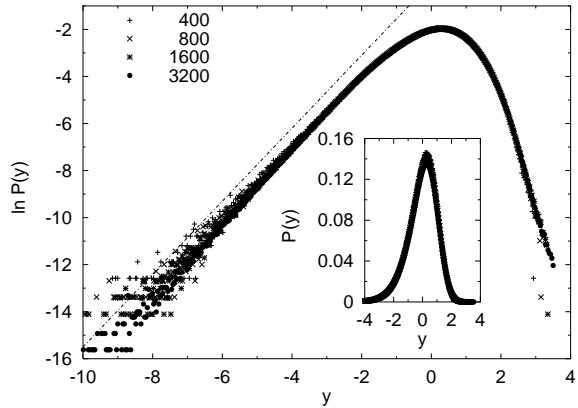}
\vglue 0.2cm
   \caption[chi]{Distribution function of the eigenvector components $P(y)$
   where $y=(\ln Q-m)/\sqrt\sigma$, $m$ is the mean and $\sigma$ is the 
   variance of $\ln Q$. The dashed--dotted line stands for the power law 
   asymptotics with a power of 1.6. The inset shows a close to Gaussian 
   form for $P(y)$.}
\label{wfprob}
\end{figure}
In Table~\ref{tab} we collected the
relevant quantites obtained from extrapolations to the thermodynamical limit.
The values of $D_2$ and $\alpha_0$ were obtained as an average of the 
corresponding exponents over the ensemble of states in the middle half of
the spectrum. The $D_2$ and $\alpha_0$ for each state were obtained using
the standard box counting algorithm~\cite{Martin}. 
\par
A further evidence of the multifractal behavior of the eigenstates is also
presented in the distribution function of the eigenvector components $P(Q)$ 
with $Q\equiv |\psi |^2$. It is broad at all length scales, i.e. a close to 
log-normal form is expected. In Fig.~\ref{wfprob} the inset shows a very nice 
normal distribution of the variable $\ln Q$ plotted after rescaling with the 
mean and the variance of it. However, the figure clearly shows deviation from
the log-normal form especially for the low-$y$ part of the distribution 
that may be described with a power law tail of the form $y^{1.6}$. 
Similar deviations have already been detected and studied 
at the quantum--Hall transition~\cite{VPJP} and are clearly seen at the 
Anderson--transition in $d=3$ systems~\cite{Varga}, as well.
\par
In summary we have investigated the spectral properties of a random
matrix ensemble with entries decaying away from the diagonal in a power--law
fashion. We found that the spacing distribution function is similar to but
deviates significantly from the semi--Poisson distribution. We have also
compared the numerically obtained functions $\Sigma^2(L)$ and $R(s)$
with the analytically known ones of the SRPM which produces semi--Poissonian
$P(s)$. These functions as well as the extrapolated value of the
level compressibility $\chi$ differ from those obtained for the SRPM.
The latter agrees well with analytical results.
We also confirm the existense of multifractal states that are
characterized by a correlation dimension $\tilde D_2$. According to our
numerical results the correlation dimension satisfies the relation 
(\ref{lcomp}) with $\chi$ obtained from the spectra.
\par
{\bf Acknowledgment:} We are indebted to F. Evers, Y. Fyodorov, M. Janssen, 
V.E. Kravtsov, A.D. Mirlin, T.H. Seligman, and P. Shukla for stimulating 
discussions and S.M. Nishigaki for providing his Mathematica code in order
to calculate the $P(s)$ for the model considered in Ref.~\cite{Nishigaki}. 
Financial support of I.V. from the Alexander von Humboldt 
Foundation and from Orsz\'agos Tudom\'anyos Kutat\'asi Alap (OTKA), 
Grant Nos. T029813, T024136 and F024135 are gratefully acknowledged.

\begin{table}[tbh]
\caption[params]{\label{tab} Relevant parameters obtained in the 
$N\to\infty$ limit from the spectral and eigenvector statistics of
the states in the middle half of the band.}
\begin{tabular}{lccc}
 & quantity   & origin                           & value           \\
\hline
 eigenvalues & $q$  & $\langle s^2\rangle^{-1}$    & $0.7342 \pm 0.0009$ \\
             & $\chi$ & $\Sigma^2(L)\to a+\chi L$  & $0.1732 \pm 0.0029$ \\
\hline
 eigenvectors & ${\tilde D}_2$ & $\langle D_2\rangle$  & $0.6617 \pm 0.0389$ \\
              & $\chi$     & $(1-{\tilde D}_2)/2$      & $0.1691 \pm 0.0195$ \\
\end{tabular}
\end{table}

\end{multicols}

\begin{references}  

\bibitem[*]{*}Permanent address: Elm\'eleti Fizika Tansz\'ek,
Budapesti M\H uszaki Egyetem, H--1521 Budapest, Hungary; 
e-mail: Imre.Varga@Physik.Uni-Marburg.De

\bibitem{Bogom}E.B. Bogomolny, U. Gerland, and C. Schmit, 
Phys. Rev. {\bf E59}, R1315 (1999), and see references therein.

\bibitem{Shkl}B.L. Shklovskii, B. Shapiro, B.R. Sears, P. Lambrianides,
and H.B. Shore,
Phys. Rev. {\bf B47}, 11487 (1993).

\bibitem{GOE}I.Kh. Zharekeshev and B.Kramer, 
Phys. Rev. Lett. {\bf 79}, 717 (1997). 

\bibitem{GUE}M.Batsch, L. Schweitzer, I.Kh. Zharekeshev, and B. Kramer, 
Phys. Rev. Lett. {\bf 77}, 1552 (1996).

\bibitem{GSE}T. Kawarabayashi, T. Ohtsuki, K. Slevin, and Y. Ono, 
Phys. Rev. Lett. {\bf 77}, 3593 (1996).

\bibitem{Daniel}D. Braun, G. Montambaux, and M. Pascaud, 
Phys. Rev. Lett. {\bf 81}, 1062 (1998).

\bibitem{Schw}L. Schweitzer and H. Potempa, 
Physica A {\bf 266}, 486 (1999).

\bibitem{Evang}G.N. Katomeris and S.N. Evangelou,
preprint {\tt cond-\-mat}/9908484.

\bibitem{IV}I. Varga, E. Hofstetter, and J. Pipek, 
Phys. Rev. Lett. {\bf 82}, 4683 (1999).

\bibitem{Mehta}M.L. Mehta, {\it Random Matrices} 
(Academic Press, Boston, 1991).

\bibitem{PhRep}T. Guhr, A. M\"uller--Groeling, and H.A. Weidenm\"uller,
Phys. Rep. {\bf 299}, 189 (1998).

\bibitem{Chalker}J.T. Chalker, I.V. Lerner, and R.A. Smith,
Phys. Rev. Lett. {\bf 77}, 554 (1996);
J.T. Chalker, V.E. Kravtsov, and I.V. Lerner, 
Pis'ma Zh. Eksp. Teor. Fiz. B. {\bf 82} 295 (1996).

\bibitem{Martin}M. Janssen, 
Phys. Rep. {\bf 295}, 1 (1998).

\bibitem{FM}Y.V. Fyodorov and A.D. Mirlin,
Phys. Rev. {\bf B55}, R16001 (1997).

\bibitem{tSel}H. Hernandez-Saldana, J. Flores, and T.H. Seligman
Phys. Rev. {\bf E60} 449 (1999).

\bibitem{Mirlin}A.D. Mirlin, Y.V. Fyodorov, F.-M. Dittes, J. Quezada,
and T.H. Seligman, 
Phys. Rev. {\bf E54} 3221 (1996).

\bibitem{Kravt}V.E. Kravtsov and K.A. Muttalib, 
Phys.~Rev.~Lett. {\bf 79} 1913 (1997).

\bibitem{Krav-loc}V.E. Kravtsov,
Ann. Phys. (Leipzig) {\bf 8} 621 (1999); V.E. Kravtsov and A.M. Tsvelik, 
preprint {\tt cond-mat}/0002120.

\bibitem{Mirlin-rev}A.D. Mirlin, 
preprint {\tt cond-mat}/9907126.

\bibitem{Evers}F. Evers and A.D. Mirlin, 
preprint {\tt cond-mat}/0001086.

\bibitem{FM-BRM}Y.V. Fyodorov and A.D. Mirlin,
Int. J. Mod. Phys. B {\bf 8}, 3795 (1994).

\bibitem{Nishigaki} S. M. Nishigaki, 
Phys. Rev. {\bf E59}, 2853 (1998)

\bibitem{Mutt}K.A. Muttalib, Y. Chen, M.E.H. Ismail, and V.N. Nicopoulos, 
Phys. Rev. Lett. {\bf 71}, 471 (1993);M. Moshe, H. Neuberger, and B. Shapiro, 
Phys. Rev. Lett. {\bf 73}, 1497 (1994). 

\bibitem{Shober}D.A. Parshin and H.R. Schober, 
Phys. Rev. Lett. {\bf 83} 4590 (1999).

\bibitem{Blecken}C. Blecken, Y. Chen, and K.A. Muttalib,
J. Phys. A: Math. Gen. {\bf 27}, L563 (1994).

\bibitem{VPJP}I. Varga, J. Pipek, M. Janssen, and K. Pratz, 
Europhys. Lett. {\bf 36} 437 (1996). 

\bibitem{Varga}I. Varga, unpublished.

\end{references}
\end{document}